\documentclass[12pt]{article}
\usepackage{epsfig,amssymb,amsmath,psfrag}
\usepackage{cite}

\def\IZ{\relax\ifmmode\mathchoice
{\hbox{\cmss Z\kern-.4em Z}}{\hbox{\cmss Z\kern-.4em Z}}
{\lower.4pt\hbox{\cmsss Z\kern-.4em Z}}
{\lower1.2pt\hbox{\cmsss Z\kern-.4em Z}}\else{\cmss Z\kern-.4em Z}\fi}
\newcommand{\Z}{\mathsf{Z}\kern -5pt \mathsf{Z}}
\def\one{{\,\hbox{1\kern-.8mm l}}}
\def\ncon{n^\Ell_{\rm con}}
\def\ncolor{n^\Ell_{\rm color}}
\def\f{\tilde{f}}
\def\Ell{{(L)}}
\def\Ellk{{(L,k)}}
\def\EllEll{{(L,L)}}
\def\EllEllmo{{(L,L-1)}}
\def\EllEllmt{{(L,L-2)}}
\def\Elltwok{{(L,2k)}}
\def\Elltwokplus{{(L,2k+1)}}
\def\Ellplus{{(L+1)}}
\def\Zero{{(0)}}
\def\One{{(1)}}
\def\Onezero{{(1,0)}}
\def\Oneone{{(1,1)}}
\def\Two{{(2)}}
\def\Twozero{{(2,0)}}
\def\Twoone{{(2,1)}}
\def\Twotwo{{(2,2)}}
\def\Three{{(3)}}
\def\Four{{(4)}}
\def\Five{{(5)}}
\def\cN {  {\cal N}  }
\def\cA {  {\cal A}  }
\def\la {\lambda}
\def\ka {\kappa}
\def\eij{e_{ij}}
\def\eonei{e_{1i}}
\def\ei{e_{i}}
\def\etwo{e_{2}}
\def\ethr{e_{3}}
\def\efour{e_{4}}
\def\efive{e_{5}}

\def \be  {\begin{equation}}
\def \ee  {\end{equation}}
\def \ba  {\begin{eqnarray}}
\def \ea  {\end{eqnarray}}

\def \Tr {\mathop{\rm Tr}\nolimits}

\def\eqn#1{eq.~(\ref{#1})} \def\Eqn#1{Equation~(\ref{#1})}
\def\eqns#1#2{eqs.~(\ref{#1}) and~(\ref{#2})}
\def\Eqns#1#2{Eqs.~(\ref{#1}) and~(\ref{#2})}

\newcommand{\nn}{\nonumber}

\begin{document}

\begin{flushright}
BOW-PH-154\\
\end{flushright}
\vspace{3mm}

\begin{center}
{\Large\bf\sf  
SU($N$) group-theory constraints  \\[1mm]
on color-ordered five-point amplitudes \\[3mm]
at all loop orders
}

\vskip 1.5cm 

Alexander C. Edison and Stephen G. Naculich\footnote{ 
Research supported in part by the National Science 
Foundation under Grant No.~PHY10-67961.}
\end{center}

\vskip 0.5cm 

\begin{center}
{\it 
Department of Physics\\
Bowdoin College\\
Brunswick, ME 04011, USA
}

\vspace{5mm}
{\tt aedison@bowdoin.edu\\
naculich@bowdoin.edu}
\end{center}

\vskip 3cm

\begin{abstract}

Color-ordered amplitudes for the scattering of $n$ particles 
in the adjoint representation of SU($N$) gauge theory 
satisfy constraints arising solely from group theory.
We derive these constraints for $n=5$ at all loop orders
using an iterative approach.
These constraints generalize well-known tree-level
and one-loop group theory relations.

\end{abstract}

\vfil\break

\section{Introduction}
\setcounter{equation}{0}

The discovery of 
the color-kinematic (BCJ) duality of gauge theory 
and the double-copy property of gravity \cite{Bern:2008qj,Bern:2010ue}
constitutes only the most recent chapter of the tremendous 
advances that have occurred in our understanding of 
perturbative gauge and gravity amplitudes over the last decade.
Tree-level relations implied by the BCJ conjecture 
have been verified in 
refs.~\cite{BjerrumBohr:2009rd,Stieberger:2009hq,Feng:2010my,Chen:2011jxa},
and the BCJ conjecture has also been tested at loop level 
for four-\cite{Bern:2010ue,Carrasco:2011hw}
and five-point \cite{Carrasco:2011mn,Bern:2011rj} amplitudes
of $\cN=4$ supersymmetric Yang-Mills theory.
This subject has been reviewed in 
refs.~\cite{Carrasco:2011hw,Sondergaard:2011iv},
which also contain references to other work on the subject.

The color structure of a gauge theory amplitude 
may be expressed by decomposing the amplitude 
in either a trace basis \cite{Bern:1990ux}
or a basis of color factors \cite{DelDuca:1999ha,DelDuca:1999rs}. 
An advantage of the trace basis decomposition
is that its coefficients, called color-ordered amplitudes,
are individually gauge-invariant.
The trace basis is also conducive to exhibiting the $1/N$ expansion 
of the gauge theory, and moreover has a close connection to the
string theory expansion of the scattering amplitudes.
Color-kinematic duality implies the existence of 
linear constraints among tree-level color-ordered amplitudes,
which were proven in 
refs.~\cite{BjerrumBohr:2009rd,Stieberger:2009hq,Feng:2010my,Chen:2011jxa}.

Even before BCJ duality is imposed, however, 
the color-ordered gauge theory amplitudes 
are known to obey various constraints 
solely as a consequence of SU($N$) group theory.
At tree level, these include the U(1) 
decoupling \cite{Green:1982sw,Mangano:1990by}
and Kleiss-Kuijf relations \cite{Kleiss:1988ne}.
These and similar group-theory relations for one-loop 
amplitudes \cite{Bern:1990ux,Bern:1994zx}
can be elegantly derived using the alternative
color decomposition of the amplitude \cite{DelDuca:1999ha,DelDuca:1999rs}.
Four-point color-ordered amplitudes 
are also known to obey group-theory relations at two loops \cite{Bern:2002tk},
and these were recently generalized to all loop orders, 
where it was shown that there exist four relations 
among color-ordered four-point amplitudes 
for each $L \ge 2$  \cite{Naculich:2011ep}.
Other recent work on constraints among loop-level amplitudes 
includes refs.~\cite{BjerrumBohr:2011xe,Feng:2011fja,Boels:2011tp,Boels:2011mn}.

The purpose of this paper is to derive 
all SU($N$) group theory relations
satisfied by five-point color-ordered amplitudes 
at two and higher loops, generalizing the
known relations at tree level and one loop. 
We employ a recursive approach \cite{Naculich:2011ep}
to derive the constraints satisfied by any $L$-loop diagram 
(containing only adjoint fields)
that can be obtained by 
%connecting any two of the external legs 
attaching a rung between two external legs
of an $(L-1)$-loop diagram. 
% with an adjoint propagator.
We assume that the most general $L$-loop color
factor can be obtained from this subset using Jacobi relations.
Then, by seeding the recursion relation with 
the six known constraints on tree-level five-point amplitudes, 
we show that there are ten constraints among 
color-ordered five-point amplitudes at each odd loop order, 
and twelve constraints at each even loop order.

In order to state our results up front,
we define $A^\Ellk$ as the part of the
$L$-loop five-point amplitude that is suppressed by $0 \le k \le L$ 
powers of $N$ relative to the leading planar amplitude,
and $A^\Ellk_\la$ as the coefficients of this amplitude
in a trace basis consisting of single-trace terms for $\la=1, \cdots, 12$
and double-trace terms for $\la=13, \cdots, 22$.
(A precise definition of this trace basis is given in the main
body of the paper.)
The six tree-level U(1) decoupling relations 
\cite{Mangano:1990by}
can be expressed as
\be
\sum_{\la=1}^{12}  A^\Zero_\la x^\Zero_{\la j}=0, 
\qquad\qquad  j = 1, \cdots 6
\ee
where $x^\Zero$ are constants defined in \eqn{defxzero}.
The ten one-loop U(1) decoupling relations \cite{Bern:1990ux}
can be expressed as
\be
A^\Oneone_\la 
= \sum_{\ka=1}^{12}  A^\Onezero_{\ka} m^\One_{\ka, \la-12}, 
\qquad\qquad \la = 13, \cdots, 22 
\ee
where $m^\One$ are constants defined in \eqn{nullonealt}.
In this paper, we show that, for all odd loop orders, 
the most-subleading-color five-point amplitudes are given by
\be
A^\EllEll_\la 
= \sum_{\ka=1}^{12}  A^\EllEllmo_{\ka} m^\One_{\ka, \la-12}, 
\qquad\qquad \la = 13, \cdots 22, \qquad {\rm  odd~}L
\label{oddloopdecoupling}
\ee
and for all even loop orders, the most-subleading-color amplitudes 
obey the six constraints\footnote{This 
relation was previously observed for $L=2$ in ref.~\cite{Feng:2011fja}.}
\be
\sum_{\la=1}^{12}  A^\EllEll_\la x^\Zero_{\la j}=0, 
\qquad\qquad  j = 1, \cdots 6, \qquad {\rm even~}L\ge 2 \,.
\label{evenloopdecoupling}
\ee
We show that five-point amplitudes obey six additional constraints 
at all even loop orders\footnote{The $L=2$ constraints 
have been independently obtained by C.~Boucher-Veronneau and L.~Dixon
\cite{BVD}.}
\be
\sum_{\la=1}^{12}  
\left( 10 A^\EllEllmt_\la x^\Zero_{\la j}
     + A^\EllEll_\la x^\Two_{\la j} \right)
+\sum_{\la=13}^{22}  A^\EllEllmo_\la x^\One_{\la-12,j}=0, \
\quad\qquad 
j = 1, \cdots 6, \qquad {\rm even~} L \ge 2
\label{newevenloop}
\ee
where $x^\One$ and $x^\Two$ are constants defined in \eqns{defxyone}{defxtwo}. 
\Eqns{evenloopdecoupling}{newevenloop} may be combined to
express each of the most-subleading-color amplitudes $A^\EllEll$ 
as linear combinations of 
$A^\EllEllmo$ and  $A^\EllEllmt$, using the constants $m^\Two$ defined
in \eqn{defmtwo}.

We have obtained the relations 
(\ref{oddloopdecoupling})--(\ref{newevenloop})
by using the connection between the color basis 
\cite{DelDuca:1999ha,DelDuca:1999rs}
and the trace basis of gauge theory amplitudes.
Since the independent color basis at $L$ loops is smaller than the trace basis,
the null eigenvectors of the transformation matrix from one basis to the other
imply constraints among the trace basis coefficients.
One can alternatively obtain some, but not all, of these constraints
by expanding the amplitude in a U($N$) trace basis, 
and observing that an amplitude containing one or more photons 
vanishes since the U(1) structure constants are 
zero \cite{Bern:1990ux,Feng:2011fja}.
Such U(1) decoupling relations can be used to derive 
\eqns{oddloopdecoupling}{evenloopdecoupling},
but not \eqn{newevenloop}.
The constraints 
(\ref{oddloopdecoupling})--(\ref{newevenloop})
reduce the number of independent $L$-loop color-ordered five-point amplitudes
from $d(L)$ to $d(L-1)$ where
$d(L) = 10 L + 2 \lfloor \frac{L}{2} \rfloor + 12$. 
No further constraints on $L$-loop color-ordered
five-point amplitudes arise from group theory alone.  

The remainder of this paper is organized as follows.
In sec.~\ref{sect-trace}, we review the trace basis
for five-point amplitudes at arbitrary loop order.
In sec.~\ref{sect-color}, we describe the color basis
and explain how each null eigenvector of the transformation matrix 
from the color basis to the trace basis implies a 
constraint among color-ordered amplitudes. 
Finally, in sec.~\ref{sect-recursive}, we utilize a recursive approach
to derive the all-loop group-theory constraints on 
five-point color-ordered amplitudes.

\section{The trace basis for all-loop five-point amplitudes}
\label{sect-trace}
\setcounter{equation}{0}

In this section, we review the trace and $1/N$ decomposition
of five-point amplitudes.
Five-point amplitudes of an SU($N$) gauge theory can be
expressed in terms of a basis $\{T_\la\}$, $\la = 1, \cdots, 22$,
of single and double traces.
We choose an explicit basis \cite{Naculich:2011fw} given by
\ba
T_1 	&=& \left[\Tr (12345) - \Tr(15432)\right],\qquad\qquad
T_7 	  = \left[\Tr (12543) - \Tr(13452)\right],\nonumber\\
T_2 	&=& \left[\Tr (14325) - \Tr(15234)\right],\qquad\qquad
T_8 	  = \left[\Tr (14523) - \Tr(13254)\right],\nonumber\\
T_3 	&=& \left[\Tr (13425) - \Tr(15243)\right],\qquad\qquad
T_9 	  = \left[\Tr (13524) - \Tr(14253)\right],\nonumber\\
T_4 	&=& \left[\Tr (12435) - \Tr(15342)\right],\qquad\qquad
T_{10}   =  \left[\Tr (12534) - \Tr(14352)\right],\nonumber\\
T_5 	&=& \left[\Tr (14235) - \Tr(15324)\right],\qquad\qquad
T_{11}   =  \left[\Tr (14532) - \Tr(12354)\right],\nonumber\\
T_6 	&=& \left[\Tr (13245) - \Tr(15423)\right],\qquad\qquad
T_{12}   =  \left[\Tr (13542) - \Tr(12453)\right],
\label{singletrace}
\ea
and 
\ba
T_{13} &=&  \Tr (12) \left[ \Tr(345) - \Tr(543) \right], \qquad\qquad
T_{18}   =  \Tr (13) \left[ \Tr(245) - \Tr(542) \right], \nonumber\\
T_{14} &=&  \Tr (23) \left[ \Tr(451) - \Tr(154) \right], \qquad\qquad
T_{19}   =  \Tr (24) \left[ \Tr(351) - \Tr(153) \right], \nonumber\\
T_{15} &=&  \Tr (34) \left[ \Tr(512) - \Tr(215) \right], \qquad\qquad
T_{20}   =  \Tr (35) \left[ \Tr(412) - \Tr(214) \right], \nonumber\\
T_{16} &=&  \Tr (45) \left[ \Tr(123) - \Tr(321) \right], \qquad\qquad
T_{21}   =  \Tr (41) \left[ \Tr(523) - \Tr(325) \right], \nonumber\\
T_{17} &=&  \Tr (51) \left[ \Tr(234) - \Tr(432) \right], \qquad\qquad
T_{22}   =  \Tr (52) \left[ \Tr(134) - \Tr(431) \right],
\label{doubletrace}
\ea
where $\Tr(12\cdots) \equiv \Tr( T^{a_1} T^{a_2} \cdots )$,
and the matrices $T^a$ are the generators in the defining representation 
of SU($N$), normalized according to $\Tr(T^a T^b) = \delta^{ab}$.
All other possible trace terms vanish in SU($N$) since $\Tr(T^a)=0$.

The $L$-loop amplitude may be further decomposed \cite{Bern:1997nh} 
in powers of $N$ as
\be
\cA^\Ell =
\sum_{\la = 1}^{12} 
\left( \sum_{k=0}^{\lfloor \frac{L}{2}  \rfloor} N^{L-2k} A^\Elltwok_\la \right) T_\la
+ \sum_{\la = 13}^{22} 
\left( \sum_{k=0}^{\lfloor \frac{L-1}{2}  \rfloor}N^{L-2k-1} A^\Elltwokplus_\la\right) T_\la
\label{Ndecomp}
\ee
where $A^{(L,0)}_\la$ are leading-color (planar) amplitudes,
and  $A^\Ellk_\la$, $k = 1, \cdots, L$,  are subleading-color amplitudes,
yielding in total $d(L)$ color-ordered amplitudes at $L$ loops, where
\be
d(L) = 
\begin{cases}
11 L + 12,  & L  {\rm ~even} \,, \\
11 L + 11,  & L  {\rm ~ odd} \,.
\end{cases}
\label{defd}
\ee
The $1/N$ expansion (\ref{Ndecomp}) suggests enlarging 
the 22-dimensional basis (\ref{singletrace}) and (\ref{doubletrace}) 
to a $d(L)$-dimensional basis which takes into account powers of $N$:
\ba
t^\Ell_{\la+ 22k} &=&   N^{L - 2k} \, T_\la  \,, 
\quad\qquad \la = 1, \cdots, 12, 
\  \qquad k = 0, \cdots, \lfloor \frac{L}{2}  \rfloor,
\nn\\ 
t^\Ell_{\la+ 22k} &=&   N^{L-2k-1} \, T_\la  \,, 
\qquad \la = 13, \cdots, 22,
\qquad k = 0, \cdots, \lfloor \frac{L-1}{2}  \rfloor \,.
\label{tracebasis}
\ea
We then write
\be
\cA^\Ell = \sum_{\la =1}^{d(L)}  A^\Ell_\la t^\Ell_\la,
\qquad
{\rm where}
\qquad
A^\Ell_{\la + 22k}=
\begin{cases}
A^\Elltwok_\la ,    & \la = 1, \cdots, 12 \,,\\
A^\Elltwokplus_\la,  & \la =13, \cdots, 22\,.
\end{cases}
\label{tracedecomp}
\ee
In the remainder of the paper, we demonstrate that
these $d(L)$ color-ordered amplitudes $A^\Ell_{\la}$
obey a set of group-theory constraints
\be
\sum_{\la=1}^{d(L)} 
A^\Ell_{\la} r^\Ell_{\la j} = 0, \qquad \qquad j = 1, \cdots, 
\begin{cases}
12,  & {\rm even~} L \ge 2\,,\\
10,  & {\rm odd~} L \,.
\end{cases}
\ee
In particular, these constraints can be used to 
express the most-subleading color amplitudes $A^\EllEll$
at each loop order as linear combinations of $A^\EllEllmo$
for $L$ odd,
and in terms of $A^\EllEllmo$ and $A^\EllEllmt$ for $L$ even.

\section{The color basis for five-point amplitudes}
\label{sect-color}
\setcounter{equation}{0}

In this section, we review the
decomposition of the amplitude in a basis
of color factors \cite{DelDuca:1999ha,DelDuca:1999rs}.
The $n$-point amplitude in a gauge theory 
containing only fields in the adjoint representation of SU($N$)
(such as pure Yang-Mills or supersymmetric Yang-Mills theory)
can be written in a loop expansion,
with the $L$-loop amplitude given by
a ``parent-graph'' decomposition \cite{Bern:2010tq}
\be
\cA^\Ell = \sum_i  a^\Ell_i c^\Ell_i \,.
\label{colordecomp}
\ee
Here $\{ c_i^\Ell \}$ represents a 
complete set\footnote{This set of color factors 
need not be independent but may satisfy constraints 
$\sum_i \ell_i c_i = 0$.
Such an overcomplete basis is usually required 
to make color-kinematic duality manifest \cite{Bern:2008qj,Carrasco:2011mn}.}
of $L$-loop $n$-point diagrams built from cubic vertices
with a factor of the SU($N$) structure constants $\f^{abc}$ at each vertex;
we have suppressed all momentum and spin dependence.    
(Contributions from Feynman diagrams containing 
quartic vertices with factors of 
$\f^{abe} \f^{cde}$,  $\f^{ace} \f^{bde}$, and $\f^{ade} \f^{bce}$
can be parceled out among other diagrams containing only cubic
vertices.)

The two decompositions (\ref{tracedecomp}) and (\ref{colordecomp})
may be related by expressing  
the color factors as linear combinations
of the trace basis (\ref{tracebasis})
\be
c_i^\Ell =  \sum_{\la=1}^{d(L)}  M^\Ell_{ i \la } t^\Ell_\la  \,.
\label{tx}
\ee
Combining \eqns{colordecomp}{tx} then yields
\be
A^\Ell_\la = \sum_i  a^\Ell_i M^\Ell_{i \la}  \,.
\label{Aa}
\ee
The number of independent color factors $\ncolor$ 
is less than the dimension of the trace basis $d(L)$, 
so the transformation matrix $M^\Ell_{i \la}$
will have a set of $\ncon$ right null 
eigenvectors\footnote{Constraints $\sum_i \ell_i c_i = 0$
among the color factors correspond to left null eigenvectors 
of the transformation matrix:  $\sum_i \ell_i M_{i \la} = 0.$}
\be
\sum_{\la=1}^{d(L)}  M^\Ell_{ i \la } r^\Ell_{\la j} =0\,, \qquad
j= 1, \cdots, \ncon,
\qquad\qquad {\rm where} \quad \ncon = d(L) - \ncolor \,.
\label{rightnull}
\ee
In other words, the vectors $r^\Ell_{\la j}$, $j = 1, \cdots, \ncon$
span the kernel 
of the transformation matrix.
\Eqn{Aa} then implies $\ncon$ constraints
on the color-ordered amplitudes 
\be
\sum_{\la=1}^{d(L)} 
A^\Ell_\la r^\Ell_{\la j} = 0, \qquad\qquad j = 1, \cdots, \ncon \,.
\label{rightrelations}
\ee
Hence, the kernel of the transformation matrix determines the 
group-theory constraints on the color-ordered amplitudes.

At tree level, we may choose an independent basis\footnote{A
larger, fifteen-dimensional basis is required to manifest 
color-kinematic duality \cite{Bern:2008qj}.}
of color factors to be \cite{DelDuca:1999ha,DelDuca:1999rs}  
\ba
&&
c^\Zero_{1} = \f^{a_1 a_2 b}\f^{b a_3 c}\f^{c a_4 a_5}\,, \hskip 0.8cm  
c^\Zero_{2} = \f^{a_1 a_4 b}\f^{b a_3 c}\f^{c a_2 a_5}\,, \hskip 0.8cm  
c^\Zero_{3} = \f^{a_1 a_3 b}\f^{b a_4 c}\f^{c a_2 a_5}\,, \nn \\
&&
c^\Zero_{4} = \f^{a_1 a_2 b}\f^{b a_4 c}\f^{c a_3 a_5}\,, \hskip 0.8cm  
c^\Zero_{5} = \f^{a_1 a_4 b}\f^{b a_2 c}\f^{c a_3 a_5}\,, \hskip 0.8cm  
c^\Zero_{6} = \f^{a_1 a_3 b}\f^{b a_2 c}\f^{c a_4 a_5}\,.
\ea
By writing 
\be
\f^{abc} = i \sqrt2 f^{abc} = \Tr( [T^a, T^b] T^c )
\ee
and using the SU($N$) identities
\ba
\Tr(P T^a) \Tr(Q T^a) &=& \Tr (PQ) - {1 \over N} \Tr(P) \Tr(Q)
\nn\\
\Tr(P T^a Q T^a) &=& \Tr (P) \Tr(Q) - {1 \over N} \Tr(P Q) 
\ea
we find that 
\be
c_i^\Zero =  \sum_{\la=1}^{12}  M^\Zero_{ i \la } t^\Zero_\la 
\qquad {\rm with} \qquad
M^\Zero
= \begin{pmatrix} \one_{6 \times 6}  & m^\Zero \end{pmatrix} 
\label{txzero}
\ee
where
\be
m^\Zero = 
\left(
\begin{array}{cccccc}
1 & 0 & 0 & 0 & 1 & 1 \\
0 & 1 & 1 & 1 & 0 & 0 \\
1 & 1 & 1 & 0 & 0 & 0 \\
0 & 0 & 0 & 1 & 1 & 1 \\
0 & 0 & 1 & 1 & 1 & 0 \\
1 & 1 & 0 & 0 & 0 & 1 \\
\end{array}
\right) \,.
\label{defmzero}
\ee
We denote the six right null eigenvectors of $M^\Zero$ by 
$x_{\la j}^\Zero$, $j=1, \cdots, 6$, where
\be
x^\Zero
= \begin{pmatrix} m^\Zero \\ - \one_{6 \times 6} \end{pmatrix}
= \left(
\begin{array}{rrrrrr}
1 & 0 & 0 & 0 & 1 & 1 \\
0 & 1 & 1 & 1 & 0 & 0 \\
1 & 1 & 1 & 0 & 0 & 0 \\
0 & 0 & 0 & 1 & 1 & 1 \\
0 & 0 & 1 & 1 & 1 & 0 \\
1 & 1 & 0 & 0 & 0 & 1 \\ [2mm]
-1& 0 & 0 & 0 & 0 & 0 \\
0 &-1 & 0 & 0 & 0 & 0 \\
0 & 0 &-1 & 0 & 0 & 0 \\
0 & 0 & 0 &-1 & 0 & 0 \\
0 & 0 & 0 & 0 &-1 & 0 \\
0 & 0 & 0 & 0 & 0 &-1
\end{array}
\right) \,.
\label{defxzero}
\ee
The six constraints on color-ordered 
tree amplitudes implied by these null vectors
\be
\sum_{\la=1}^{12}  A^\Zero_\la x^\Zero_{\la j}=0, 
\qquad\qquad  j = 1, \cdots 6
\label{treeleveldecoupling}
\ee
are precisely the tree-level U(1) decoupling relations for
five-point amplitudes \cite{Mangano:1990by}.
In the next section, we will use \eqns{txzero}{defmzero} as the
starting point of a recursive procedure to derive 
higher-loop relations among five-point amplitudes.

\section{Relations among $L$-loop five-point amplitudes}
\label{sect-recursive}
\setcounter{equation}{0}

In this section, we employ a recursive procedure \cite{Naculich:2011ep}
to obtain the set of right null eigenvectors at 
arbitrary loop level,
which then determine the constraints 
on the color-ordered amplitudes.
An $(L+1)$-loop diagram may be obtained from an $L$-loop diagram by
%connecting 
attaching a rung between 
two of its external legs, $i$ and $j$. 
%with an adjoint propagator.
This corresponds to contracting its color factor with
$\eij \f^{a_i a'_i b} \f^{b a'_j a_j}$.
If $i$ and $j$ are not adjacent, this will convert a planar diagram into
a nonplanar diagram.

First consider the effect of this procedure on the trace basis 
(\ref{singletrace})-(\ref{doubletrace})
\be 
T_\la \longrightarrow \sum_{\ka=1}^{22} G_{\la\ka} T_\ka 
\, , \qquad\qquad {\rm with} \qquad\qquad
G = 
\begin{pmatrix} 
N A & B \\ C & N D
\end{pmatrix}
\label{defG}
\ee
where explicit expressions for the submatrices $A$, $B$, $C$, and $D$
are given in the appendix.
On the expanded basis (\ref{tracebasis}),
the same procedure yields
\be
t^\Ell_\la \to \sum_{\ka=1}^{d(L+1)} g_{\la\ka} t^\Ellplus_\ka 
\ee
where $g^\Ell$ is the $d(L) \times d(L+1)$ matrix
\be
g^\Ell = \left( \begin{array}{cccccc}
	A & B & 0 & 0 & 0 & \hdots \\
	0 & D & C & 0 & 0 & \hdots \\
	0 & 0 & A & B & 0 & \hdots \\
	0 & 0 & 0 & D & C & \hdots \\
	\vdots & \vdots & \vdots & \vdots & \vdots & \ddots \\
	\end{array} \right) \,.
\ee

Now consider a complete set $\{ c_i^\Ell \}$ of $L$-loop
color factors, expressed in terms of the trace basis via \eqn{tx}.
%Connecting 
Attaching a rung between two external legs of $c_i^\Ell$ 
%with an adjoint propagator 
yields
\be
c_i^\Ell \to   \sum_{\la=1}^{d(L)}  \sum_{\ka=1}^{d(L+1)} 
M^\Ell_{ i \la } g^\Ell_{\la\ka} t^\Ellplus_\ka  \,.
\ee
If we assume that this procedure generates a complete set of 
$(L+1)$-loop color factors $\{ c_i^\Ellplus \}$,
then the kernel of $M^\Ellplus_{i \ka}$  
is the kernel of $\sum_{\la=1}^{d(L)}  M^\Ell_{ i \la } g^\Ell_{\la\ka}$; 
in other words,
\be
\sum_{\la=1}^{d(L)}   \sum_{\ka=1}^{d(L+1)} M^\Ell_{i  \la} g^\Ell_{\la \ka} r^\Ellplus_{\ka j} = 0 \,.
\label{recursion}
\ee
The solutions $r^\Ellplus_{\ka j}$ of this equation are given by the solutions of 
\be
\sum_{\ka=1}^{d(L+1)}  g^\Ell_{\la \ka} r^\Ellplus_{\ka j} = {\rm linear~combination~of~} \{ r^\Ell_{\la j'} \} 
\label{recursive} 
\ee
by virtue of \eqn{rightnull}.
We stress that \eqns{recursion}{recursive} 
must hold for arbitrary values of the parameters
$\eij$ in the matrices (\ref{defABCD}).

Our task is now to solve \eqn{recursive} recursively. 
We begin with the tree-level 
transformation matrix $M^\Zero$ given in \eqns{txzero}{defmzero}. 
At one loop, there are ten solutions of 
\be
\sum_{\la=1}^{12}   \sum_{\ka=1}^{22} 
M^\Zero_{i  \la} g^\Zero_{\la \ka} r^\One_{\ka j} = 0
\label{onelooprecursion}
\ee
where $g^\Zero = (A \quad B)$;
these null eigenvectors are given by 
\be
r^\One =  \left( 
                 \begin{array}{cc}
                  10 x^\Zero  &  0        \\
                  x^\One   &  y^\One  
                 \end{array}
                 \right)
\label{nullone}
\ee
where $x^\Zero$  was defined in \eqn{defxzero}, and
$x^\One$ and $y^\One$  are given by\footnote{Each
$x^\One_j$ is defined only up to the addition of arbitrary linear 
combinations of the $y_j^\One$.
This freedom could be used, for example, 
to set the bottom four entries of $x_j^\One$ to zero.
We have chosen the $x_j^\One$ to be consistent with the two-loop eigenvectors
given below.} 
\be
x^\One =  \left(
                 \begin{array}{rrrrrr}
                   -2 & 0 & 0 & 2 & 1 & -1 \\
                   -2 & -1 & -2 & -2 & -1 & -2 \\
                   -1 & -2 & -2 & -1 & -2 & -2 \\
                   1 & 2 & 0 & 0 & -2 & -1 \\
                   -1 & 1 & -1 & 1 & -1 & 1 \\
                   -1 & 1 & 2 & 0 & 0 & -2 \\
                   -2 & -2 & -1 & -2 & -2 & -1 \\
                   0 & 0 & 2 & 1 & -1 & -2 \\
                   0 & 2 & 1 & -1 & -2 & 0 \\
                   -2 & -1 & 1 & 2 & 0 & 0 \\
              \end{array}
             \right),
%                 5 \left(\begin{array}{rrrrrrrrrr}
%                  0 & 0 & 0 & 1 & 1 & 0 \\
%                  0 & 0 & -1 & -1 & 0 & 0 \\
%                  -1 & -1 & 0 & 0 & -1 & -1 \\
%                  1 & 1 & 0 & 0 & 0 & 0 \\
%                  -1 & 0 & 0 & 0 & -1 & 0 \\
%                  -1 & 0 & 1 & 0 & -1 & -1 \\
%                  0 & 0 & 0 & 0 & 0 & 0 \\
%                  0 & 0 & 0 & 0 & 0 & 0 \\
%                  0 & 0 & 0 & 0 & 0 & 0 \\
%                  0 & 0 & 0 & 0 & 0 & 0
%                 \end{array} \right)
\qquad\qquad 
y^\One  = \left(
                 \begin{array}{rrrr}
		   1 & 0 & 1 & 0 \\
                  0 & 1 & 0 & 1 \\
                  -1 & -1 & 0 & -1 \\
                  1 & 0 & 0 & 1 \\
                  -1 & 0 & -1 & -1 \\
                  -1 & -1 & -1 & -1 \\
                  1 & 0 & 0 & 0 \\
                  0 & 1 & 0 & 0 \\
                  0 & 0 & 1 & 0 \\
                  0 & 0 & 0 & 1
                 \end{array}
                 \right) \,.
\label{defxyone}
\ee
The last four eigenvectors of \eqn{nullone} imply constraints
among the one-loop double-trace coefficients
\be
\sum_{\la=13}^{22}  A^\Oneone_\la y^\One_{\la-12,j}=0, 
\qquad\qquad  j = 1, \cdots 4
\ee
while the other six eigenvectors imply relations
between the one-loop single-trace and double-trace coefficients.
Equivalently we can write the ten one-loop null eigenvectors \eqn{nullone} as
\be
r^\One
= \begin{pmatrix}  m^\One \\ - \one_{10 \times 10} \end{pmatrix},
\qquad
{\rm where}
\quad
m^\One =\left(
	\begin{array}{rrrrrrrrrr}
                   1 & 1 & 1 & 1 & 1 & 1 & 1 & 1 & 1 & 1 \\
                   -1 & 1 & 1 & -1 & -1 & -1 & 1 & -1 & -1 & -1 \\
                   1 & 1 & 1 & -1 & 1 & -1 & 1 & -1 & -1 & 1 \\
                   -1 & 1 & 1 & 1 & -1 & 1 & 1 & 1 & 1 & -1 \\
                   -1 & 1 & 1 & 1 & 1 & -1 & 1 & -1 & 1 & -1 \\
                   1 & 1 & 1 & -1 & -1 & 1 & 1 & 1 & -1 & 1 \\
                   -1 & -1 & 1 & 1 & -1 & -1 & -1 & 1 & -1 & -1 \\
                   1 & 1 & -1 & 1 & 1 & 1 & -1 & -1 & 1 & -1 \\
                   -1 & -1 & -1 & -1 & -1 & 1 & 1 & 1 & 1 & 1 \\
                   1 & -1 & 1 & 1 & 1 & -1 & -1 & 1 & -1 & 1 \\
                   1 & 1 & -1 & -1 & -1 & 1 & -1 & -1 & -1 & -1 \\
                   -1 & -1 & -1 & -1 & 1 & -1 & 1 & -1 & 1 & 1 \\
                  \end{array}
            \right) \,.
\label{nullonealt}
\ee
These eigenvectors allow us to express each of 
$A^\One_{13}$ through $A^\One_{22}$
(which correspond to double-trace coefficients)
in terms of a linear combination of 
$A^\One_{1}$ through $A^\One_{12}$
(which correspond to single-trace coefficients)
namely
\be
A^\Oneone_\la 
= \sum_{\ka=1}^{12}  A^\Onezero_{\ka} m^\One_{\ka, \la-12}, 
\qquad\qquad \la = 13, \cdots, 22  \,.
\label{oneloopdecoupling}
\ee
These are precisely the one-loop 
U(1) decoupling relations \cite{Bern:1990ux}
as previously observed \cite{Naculich:2011fw}.

Next we turn to the two-loop case.
The two-loop null eigenvectors $r^\Two$ satisfy
\be
\sum_{\la=1}^{22}   \sum_{\ka=1}^{34} 
M^\One_{i  \la} g^\One_{\la \ka} r^\Two_{\ka j} = 0,
\qquad {\rm where} \qquad
g^\One = \left( \begin{array}{ccc}
	A & B & 0 \\
	0 & D & C 
		\end{array} \right) \,.
\label{twolooprecursion}
\ee
In ref.~\cite{Naculich:2011fw}, the form of $M^\One$ was obtained by
expressing the independent twelve-dimensional basis 
of pentagon color factors in terms of the one-loop trace basis.
In this paper, we instead (and equivalently) construct $M^\One$
as the matrix that annihilates the set of
known one-loop null eigenvectors (\ref{nullonealt}),
namely,
\be
M^\One
= \begin{pmatrix} \one_{12 \times 12}  & m^\One \end{pmatrix} \,.
\ee
Then one finds that \eqn{twolooprecursion} has twelve solutions 
\be
r^\Two =  \left( 
                 \begin{array}{cc}
                  10 x^\Zero  &  0        \\
                  x^\One   &  0        \\
                  x^\Two   &  x^\Zero  \\
                 \end{array}
                 \right) \,,
\label{nulltwo} 
\ee
where
$x^\Zero$ and $x^\One$ were given in \eqns{defxzero}{defxyone},
and  $x^\Two$ is given by
\be
x^\Two =  \left(
                 \begin{array}{rrrrrr}
                   1 & 2 & 4 & 2 & 1 & 0 \\
                   2 & 1 & 0 & 1 & 2 & 4 \\
                   1 & 0 & 1 & 2 & 4 & 2 \\
                   2 & 4 & 2 & 1 & 0 & 1 \\
                   4 & 2 & 1 & 0 & 1 & 2 \\
                   0 & 1 & 2 & 4 & 2 & 1 \\ [2mm]
                   0 & 0 & 0 & 0 & 0 & 0 \\
                   0 & 0 & 0 & 0 & 0 & 0 \\
                   0 & 0 & 0 & 0 & 0 & 0 \\
                   0 & 0 & 0 & 0 & 0 & 0 \\
                   0 & 0 & 0 & 0 & 0 & 0 \\
                   0 & 0 & 0 & 0 & 0 & 0
                  \end{array}
\right) \,.
\label{defxtwo}
\ee
We observe that the last six eigenvectors of \eqn{nulltwo} 
imply constraints
among $A^\Ell_{23}$ through $A^\Ell_{34}$, 
that is, among the two-loop subleading-color single-trace coefficients:
\be
\sum_{\la=1}^{12}  A^\Twotwo_\la x^\Zero_{\la j}=0,
\qquad\qquad  j = 1, \cdots 6 \,.
\label{twoloopdecoupling}
\ee
These are simply the two-loop U(1) decoupling relations,
analogous to the tree-level U(1) decoupling relations
(\ref{treeleveldecoupling}),
and can be obtained by an extension of the arguments 
of ref.~\cite{Bern:1990ux}.
\Eqn{twoloopdecoupling}
was previously observed in ref.~\cite{Feng:2011fja}.
In that paper, it was also shown that analogs of the Kleiss-Kuijf relations
hold for the two-loop subleading-color single-trace coefficients
through seven points, but are modified for eight-point functions. 

The first six eigenvectors  of \eqn{nulltwo} 
correspond to additional relations among 
two-loop leading-color and subleading-color 
five-point amplitudes\footnote{C.~Boucher-Veronneau and 
L.~Dixon have independently obtained these relations 
by recasting an independent two-loop color basis into the trace basis 
\cite{BVD}.}
\be
\sum_{\la=1}^{12}  
\left( 10 A^\Twozero_\la x^\Zero_{\la j}
     + A^\Twotwo_\la x^\Two_{\la j} \right)
+\sum_{\la=13}^{22}  A^\Twoone_\la x^\One_{\la-12,j}=0, 
\qquad\qquad 
  j = 1, \cdots 6 \,.
\ee
These constraints, however, cannot be obtained by U(1) decoupling
arguments.

The twelve two-loop null eigenvectors (\ref{nulltwo}) 
can equivalently be expressed as
\be
r^\Two
= \begin{pmatrix} m^\Two\\ 
                  -2 \!\times\!\! \one_{12 \times 12} 
  \end{pmatrix}
\label{nulltwoalt}
\ee
where
\be
m^\Two = \left(
\begin{array}{rrrrrrrrrrrr}
   		  0 & -2 & -4 & 2 & -4 & 2 & -2 & -4 & -10 & -4 & -2 & 4 \\
                  -2 & 0 & 2 & -4 & 2 & -4 & -4 & -2 & 4 & -2 & -4 & -10 \\
                  -4 & 2 & 0 & -2 & -4 & 2 & -2 & 4 & -2 & -4 & -10 & -4 \\
                  2 & -4 & -2 & 0 & 2 & -4 & -4 & -10 & -4 & -2 & 4 & -2 \\
                  -4 & 2 & -4 & 2 & 0 & -2 & -10 & -4 & -2 & 4 & -2 & -4 \\
                  2 & -4 & 2 & -4 & -2 & 0 & 4 & -2 & -4 & -10 & -4 & -2 \\
                  -2 & -4 & -2 & -4 & 10 & 4 & 0 & -2 & 4 & 2 & 4 & -2 \\
                  -4 & -2 & 4 & 10 & -4 & -2 & -2 & 0 & -2 & 4 & 2 & 4 \\
                  10 & 4 & -2 & -4 & -2 & -4 & 4 & -2 & 0 & -2 & 4 & 2 \\
                  -4 & -2 & -4 & -2 & 4 & 10 & 2 & 4 & -2 & 0 & -2 & 4 \\
                  -2 & -4 & 10 & 4 & -2 & -4 & 4 & 2 & 4 & -2 & 0 & -2 \\
                  4 & 10 & -4 & -2 & -4 & -2 & -2 & 4 & 2 & 4 & -2 & 0 \\ [2mm]
                  1 & 1 & -1 & -1 & 1 & -1 & -1 & -1 & 1 & 1 & 1 & -1 \\
                  1 & 1 & -1 & -1 & 1 & 1 & 1 & 1 & 1 & 1 & 1 & 1 \\
                  1 & 1 & 1 & 1 & -1 & -1 & 1 & 1 & 1 & 1 & 1 & 1 \\
                  1 & 1 & 1 & -1 & -1 & -1 & 1 & 1 & 1 & -1 & -1 & -1 \\
                  1 & -1 & 1 & -1 & 1 & -1 & 1 & -1 & 1 & -1 & 1 & -1 \\
                  -1 & 1 & -1 & -1 & 1 & 1 & -1 & 1 & 1 & 1 & -1 & -1 \\
                  -1 & -1 & 1 & 1 & 1 & 1 & 1 & 1 & 1 & 1 & 1 & 1 \\
                  -1 & 1 & 1 & 1 & -1 & -1 & -1 & 1 & 1 & 1 & -1 & -1 \\
                  -1 & -1 & 1 & -1 & 1 & 1 & 1 & 1 & 1 & -1 & -1 & -1 \\
                  -1 & -1 & 1 & 1 & 1 & -1 & -1 & -1 & 1 & 1 & 1 & -1 \\
                 \end{array}
                 \right) \,.
\label{defmtwo}
\ee
These allow one to express all of the 
most-subleading-color amplitudes $A^\Twotwo$
in terms of linear combinations of $A^\Twozero$ and $A^\Twoone$.

Three-loop eigenvectors are the solutions of
\be
\sum_{\la=1}^{34}   \sum_{\ka=1}^{44} 
M^\Two_{i  \la} g^\Two_{\la \ka} r^\Three_{\ka j} = 0 \,.
\label{threelooprecursion}
\ee
One could construct $M^\Two$ from an independent
basis of two-loop color factors, 
but instead we simply write it as the matrix whose kernel is 
given by \eqn{nulltwoalt}, 
namely
\be
M^\Two_{i\la} 
= \begin{pmatrix} 2\! \times\!\!\one_{22 \times 22}  & m^\Two \end{pmatrix} \,.
\ee
One then finds that \eqn{threelooprecursion} has precisely
ten solutions,
\be
r^\Three =  
\left( 
\begin{array}{rr}
0 & 0        \\
0 & 0        \\
10 x^\Zero  &  0        \\
x^\One   &  y^\One  
\end{array}
\right)
\label{nullthree}
\ee
where $x^\Zero$, $x^\One$, and $y^\One$ were given 
in \eqns{defxzero}{defxyone}.
These are exactly analogous to the one-loop null eigenvectors (\ref{nullone}).

The fact that there are no three-loop null eigenvectors
with non-zero entries in the top two blocks makes it
possible to solve the recursion relations to all loop orders.
This is based on the following observation.
We have written the $L$-loop null eigenvectors in block form,
with $r^\Ell$ having $(L+1)$ blocks of alternating height 12 and 10.
By setting $\etwo=\ethr=\efour=\efive=1$ 
in the matrices (\ref{defABCD}),
one obtains $A = D = \one$ and $B = C = 0$.
Then \eqn{recursive} implies that the top $L$ blocks 
of $r^\Ell$ must be a linear combination of 
$(L-1)$-loop null eigenvectors.
Then, since the top two blocks of $r^\Three$ vanish,
the top two blocks of $r^\Four$ must also vanish,
so that \eqn{recursive} for $r^\Four$ 
is precisely equivalent to \eqn{recursive} for $r^\Two$.
Similarly, \eqn{recursive} for $r^\Five$ 
is precisely equivalent to \eqn{recursive} for $r^\Three$, 
and so forth.
Thus, there are ten null eigenvectors at each odd-loop order
and twelve null eigenvectors at each even-loop order 
(except at tree level, where there are six),
which are given explicitly by 
\be
r^{(2\ell+1)}=  
\left( 
\begin{array}{rr}
\vdots & \vdots \\
0   &   0    \\
0   &   0    \\
10 x^\Zero  &  0        \\
x^\One   &  y^\One  
\end{array}
\right), 
\qquad
r^{(2\ell+2)}=  
\left( 
\begin{array}{rr}
\vdots & \vdots \\
0   &   0    \\
10 x^\Zero  &  0        \\
x^\One   &  0        \\
x^\Two   &  x^\Zero  \\
\end{array}
\right) \,.
\ee
Finally, the $L$-loop relations among color-ordered five-point amplitudes
implied by these null eigenvectors 
are given in 
eqs.~(\ref{oddloopdecoupling})--(\ref{newevenloop}).

\section{Conclusions}
\setcounter{equation}{0}

In this paper, we have derived group-theory identities 
for color-ordered five-point amplitudes 
in SU($N$) gauge theories at all loop orders.
We used a recursive procedure 
to derive the constraints on $L$-loop color factors 
that are generated by 
%connecting 
attaching a rung between 
two external legs of an $(L-1)$-loop color factor.
Assuming that all $L$-loop color factors are linear
combinations of those just described (i.e., via Jacobi 
relations), the constraints derived 
apply to all $L$-loop color-ordered amplitudes.

At odd-loop levels, we obtained ten relations,
analogous to the one-loop U(1) decoupling relations,
which allow one to express the most-subleading-color amplitudes
$A^\EllEll$ in terms of the second-most-subleading-color
amplitudes $A^\EllEllmo$.
At even-loop levels, we obtained six relations
that relate the most-subleading-color amplitudes $A^\EllEll$
to one another,
analogous to the tree-level U(1) decoupling relations.
For even $L \ge 2$, we obtained six additional 
relations relating $A^\EllEll $ to other amplitudes,
which cannot be obtained by U(1) decoupling.
All twelve of the even-loop relations can be combined to express
$A^\EllEll$ as linear combinations 
of $A^\EllEllmo$ and $A^\EllEllmt$.

It would clearly be of interest to extend these results
to six-point amplitudes and beyond. 

\section*{Acknowledgments}

It is a pleasure to thank C.~Boucher-Veronneau for 
discussions and for useful comments on the manuscript.

\vfil\break

\appendix 
\section{Appendix} 
\setcounter{equation}{0}

The submatrices that comprise $G$ in \eqn{defG} are given by
\ba
A &=&  \left(
\begin{smallmatrix}
   \etwo + \efive& 0& 0& 0& 0& 0& 0& 0& 0& 0& 0& 0\\
   0& \efour + \efive& 0& 0& 0& 0& 0& 0& 0& 0& 0& 0\\
    0& 0& \ethr + \efive& 0& 0& 0& 0& 0& 0& 0& 0& 0\\
    0& 0& 0& \etwo + \efive& 0& 0& 0& 0& 0& 0& 0& 0\\
    0& 0& 0& 0& \efour + \efive& 0& 0& 0& 0& 0& 0& 0\\
    0& 0& 0& 0& 0& \ethr + \efive& 0& 0& 0& 0& 0& 0\\
    0& 0& 0& 0& 0& 0& \etwo + \ethr& 0& 0& 0& 0& 0\\
    0& 0& 0& 0& 0& 0& 0& \ethr + \efour& 0& 0& 0& 0\\
    0& 0& 0& 0& 0& 0& 0& 0& \ethr + \efour& 0& 0& 0\\
    0& 0& 0& 0& 0& 0& 0& 0& 0& \etwo + \efour& 0& 0\\
    0& 0& 0& 0& 0& 0& 0& 0& 0& 0& \etwo + \efour& 0\\
    0& 0& 0& 0& 0& 0& 0& 0& 0& 0& 0& \etwo + \ethr
\end{smallmatrix} 
\right), 
\nn \\ [5mm]
B &=& \left(
\begin{smallmatrix}
\etwo - \ethr& -\ethr + \efour& 0& \ethr - \efour& -\efour + \efive& 0& 0& 0& 0& 0\\
0& 0& \etwo - \ethr& 0& \etwo - \efive& 0& 0& 0& \ethr - \efour& \etwo - \ethr\\
0& 0& \etwo - \efour& 0& -\etwo + \efive& -\ethr + \efour& 0& 0& 0& -\etwo + \efour \\ 
-\etwo + \efour& 0& 0& 0& \ethr - \efive& 0& \ethr - \efour& -\ethr + \efour& 0& 0 \\
0& 0& 0& 0& -\ethr + \efive& 0& -\etwo + \ethr& -\etwo + \ethr& -\etwo + \efour& 0 \\
0& -\etwo + \efour& 0& -\etwo + \efour& \efour - \efive& -\etwo + \ethr& 0& 0& 0& 0 \\
-\etwo + \efive& 0& -\efour + \efive& 0& 0& -\ethr + \efour& 0& 0& 0& -\efour + \efive\\
 0& -\etwo + \efive& 0& \etwo - \efive& 0& -\etwo + \ethr& 0& 0& \efour - \efive& 0 \\
    0& 0& 0& 0& 0& \ethr - \efive& -\etwo + \efive& \etwo - \efive& -\etwo + \efour& 0 \\
    \etwo - \efive& 0& -\ethr + \efive& 0& 0& 0& 0& 0& \ethr - \efour& \ethr - \efive\\
    \etwo - \ethr& -\ethr + \efive& 0& -\ethr + \efive& 0& 0& 0& 0& -\efour + \efive& 0 \\
    -\etwo + \efour& 0& 0& 0& 0& -\ethr + \efive& -\efour + \efive& -\efour + \efive& 0& 0
\end{smallmatrix}
\right),
\nn \\ [5mm]
C & = & \left(
\begin{smallmatrix}
  -\ethr + \efive& 0& 0& \efour - \efive& 0& 0&
            -\ethr + \efive& 0& 0& \efour - \efive&  -\ethr + \efour& -\ethr + \efour\\
      \etwo - \ethr& 0& 0& 0& 0& -\etwo + \ethr&
            0& -\etwo + \ethr& 0& 0& \etwo - \ethr& 0 \\
    0& -\ethr + \efour& \ethr - \efour& 0& 0& 0&
            \ethr - \efour& 0& 0& -\ethr + \efour& 0& 0 \\
    -\efour + \efive& 0& 0& 0& 0& \efour - \efive&
            0& \efour - \efive& 0& 0& -\efour + \efive& 0 \\
    \etwo - \efour& \etwo - \efour& -\etwo + \ethr& -\etwo + \ethr& -\ethr + \efour&
            -\ethr + \efour&  0& 0& 0& 0& 0& 0 \\
    0& 0& \efour - \efive& 0& 0& -\etwo + \efive&
           -\etwo + \efour& -\etwo + \efour& \efour - \efive& 0& 0& -\etwo + \efive\\
    0& 0& 0& \etwo - \efour& -\etwo + \efour& 0&
           0& 0& -\etwo + \efour& 0& 0& \etwo - \efour\\
    0& 0& 0& -\ethr + \efive& \ethr - \efive& 0&
            0& 0& \ethr - \efive& 0& 0& -\ethr + \efive\\
    0& \ethr - \efive& 0& 0& -\etwo + \efive& 0&
           0& \ethr - \efive& -\etwo + \ethr&  -\etwo + \ethr& -\etwo + \efive& 0 \\
    0& \etwo - \efive& -\etwo + \efive& 0& 0& 0&
          -\etwo + \efive& 0& 0& \etwo - \efive& 0& 0
\end{smallmatrix}
\right),
\nn  \\ [5mm]
D &=&
\left(\begin{smallmatrix}
    2 \etwo& 0& 0& 0& 0& 0& 0& 0& 0& 0 \\
    0& \efour + \efive& 0& 0& 0& 0& 0& 0& 0& 0 \\
    0& 0& \etwo + \efive& 0& 0& 0& 0& 0& 0& 0 \\
    0& 0& 0& \etwo + \ethr& 0& 0& 0& 0& 0& 0 \\
    0& 0& 0& 0& 2 \efive& 0& 0& 0& 0& 0 \\
    0& 0& 0& 0& 0& 2 \ethr& 0& 0& 0& 0 \\
    0& 0& 0& 0& 0& 0& \ethr + \efive& 0& 0& 0 \\
    0& 0& 0& 0& 0& 0& 0& \etwo + \efour& 0& 0 \\
    0& 0& 0& 0& 0& 0& 0& 0& 2 \efour& 0 \\
    0& 0& 0& 0& 0& 0& 0& 0& 0& \ethr + \efour
\end{smallmatrix}
\right)
\label{defABCD}
\ea
where the coefficient of $\ei \equiv \eonei$ denotes the matrix obtained
by 
%connecting 
attaching a rung between external legs 1 and $i$ of the trace basis
(\ref{singletrace}), (\ref{doubletrace}).
% with an adjoint propagator.  
We have omitted 
for the sake of readability
terms in these matrices that
correspond to connecting other pairs of external legs.
They may be obtained 
from these matrices via permutations of rows and columns.

\vfil\break

\end{document}